\documentclass[aps,prd,unsortedaddress,superscriptaddress,showpacs,nofootinbib,twocolumn,10pt,floatfix]{revtex4-2} %notitlepage,
\usepackage{natbib}
\usepackage[utf8]{inputenc}
\usepackage{newunicodechar}
\usepackage{array}
\usepackage{multirow}
\usepackage{graphicx,color}
\usepackage{relsize}
\usepackage{slashed}
\usepackage[normalem]{ulem}
\usepackage{tabu}
\usepackage{rotating}
\usepackage{bigstrut}
\usepackage{makecell}
\usepackage[dvipsnames]{xcolor}
\usepackage[colorlinks=true, linkcolor=blue, citecolor=Green, urlcolor=blue]{hyperref}
\usepackage{amsmath}
\usepackage{amssymb,bm}
\usepackage{amsthm}
\usepackage{mathrsfs}
\usepackage{array}
\usepackage[all]{xy}
\usepackage{euscript}
\usepackage{enumerate}
\usepackage{mathtools}
\usepackage{soul}
\usepackage{orcidlink}

\allowdisplaybreaks

\newcommand{\bonn}{\affiliation{Helmholtz Institut f\"{u}r Strahlen- und Kernphysik, Bethe Center for Theoretical Physics
and Cluster of Excellence  ``Color meets Flavor'', Universit\"{a}t Bonn, D-53115 Bonn, Germany}}
\newcommand{\julich}{\affiliation{Institute for Advanced Simulation (IAS-4) and Cluster of Excellence ``Color meets Flavor'', Forschungszentrum J\"ulich, D-52425 J\"ulich, Germany}}
\newcommand{\peng}{\affiliation{Peng Huanwu Collaborative Center for Research and Education, International Institute for Interdisciplinary and Frontiers, Beihang University, Beijing 100191, China}}

\newcommand{\NNbar}{N\bar N}
\newcommand{\ppbar}{p\bar p}
\newcommand{\nnbar}{n\bar n}
\newcommand{\ee}{e^+e^-}

\newcommand{\chisqdof}{\chi^2/\mathrm{dof}}

\begin{document}

\title{Understanding the near-threshold structures in $\ee$ annihilation\\ from a unified $\NNbar$-interaction perspective}

\author{Teng Ji\orcidlink{0000-0003-0366-1042}}
\email{teng@hiskp.uni-bonn.de}
\bonn

\author{Ulf-G. Mei{\ss}ner\orcidlink{0000-0003-1254-442X}}\email{meissner@hiskp.uni-bonn.de}
\bonn\julich\peng

% \date{}

\begin{abstract}
Near-threshold structures have been observed in the cross sections for $\ee\to\ppbar$, $\ee\to\nnbar$, and several non-baryonic final states in the vicinity of the $\NNbar$ thresholds. We investigate whether these structures can be understood as manifestations of a common $\NNbar$ final-state interaction. The strong $\NNbar$ interaction is taken from the chiral EFT description of the coupled ${}^3S_1$-${}^3D_1$ system constrained by low-energy $\NNbar$ scattering data. With this interaction fixed, the $\ppbar$ and $\nnbar$ cross sections are described by fitting only short-distance electromagnetic production sources, which are assumed to vary slowly over the near-threshold region. The resulting $\NNbar$ production amplitudes are then used as input for five inelastic hadronic channels. A simultaneous description of the near-threshold cross sections is obtained, indicating that the observed structures can be consistently interpreted as consequences of the same underlying $\NNbar$ dynamics, without introducing separate narrow resonances in individual channels.
\end{abstract}

\maketitle

\section{Introduction}
\label{sec:intro}
Near-threshold structures associated with the nucleon-antinucleon thresholds have been observed in a variety of processes and have become a recurrent topic in hadron spectroscopy.
In $J/\psi$ decays, pronounced enhancements close to the $\ppbar$ threshold have been reported in the $\ppbar$ invariant-mass spectrum~\cite{BES:2003aic,CLEO:2010fre,BESIII:2010vwa,BESIII:2011aa}, together with several other threshold structures in light-meson production channels~\cite{BES:2005ega,BESIII:2010gmv,BESIII:2013sbm,BESIII:2016fbr,BESIII:2023vvr}.
Similar enhancements in the $\ppbar$ invariant-mass distribution have also been observed in $B$ decays~\cite{Belle:2002fay,Belle:2002bro,Belle:2007oni,LHCb:2013njz,LHCb:2014nix}.
In the timelike electromagnetic sector, the effective form factors of the proton and neutron show strong rises close to the corresponding thresholds~\cite{BaBar:2013ves,CMD-3:2015fvi,CMD-3:2018kql,BESIII:2021rqk,SND:2022wdb}.
Moreover, nontrivial structures are visible near the $\NNbar$ thresholds in a number of hadronic cross sections measured in $e^+e^-$ annihilation, including
$3(\pi^+\pi^-)$, $2(\pi^+\pi^-\pi^0)$, $2(\pi^+\pi^-)\pi^0$, $\omega\pi^+\pi^-\pi^0$, and $K^+K^-\pi^+\pi^-$~\cite{BaBar:2006vzy,BaBar:2007ptr,CMD-3:2013nph,Lukin:2015vsa,CMD-3:2018kql}.
The occurrence of such structures in both baryonic and non-baryonic final states suggests that the opening of the $\ppbar$ and $\nnbar$ channels may leave observable imprints beyond direct baryon-pair production. This is a generic possibility for near-threshold phenomena, as required by unitarity and analyticity~\cite{Guo:2017jvc,Dong:2020hxe,Zhang:2024qkg}.

This observation raises a natural question: can the near-threshold structures seen in different channels be understood as manifestations of a common $\NNbar$ dynamics?
One particularly relevant mechanism is that the opening of the $\ppbar$ and $\nnbar$ channels generates nonanalytic threshold effects, which are then transmitted to other hadronic final states through the strong interaction.
% In this case, the small proton-neutron mass difference, the Coulomb interaction in the charged channel, and the coupled $\ppbar$-$\nnbar$ dynamics can all affect the detailed line shapes in the immediate vicinity of the threshold.

Several theoretical explanations have been proposed for some of these near-threshold structures.
One class of interpretations attributes the observed anomalies to near-threshold baryonium states~\cite{Datta:2003iy,Ding:2005gh,Wang:2006sna,Dedonder:2009bk} or to $\NNbar$ final-state interactions~\cite{Zou:2003zn,Sibirtsev:2004id,Haidenbauer:2006au,Haidenbauer:2006dm,Chen:2008ee,Chen:2010an,Chen:2011yu,Zhou:2012ui,Haidenbauer:2014kja,Kang:2015yka,Haidenbauer:2015yka,Dmitriev:2015qyt,Milstein:2018orb,Milstein:2022tfx,Lin:2021umz,Yang:2022kpm,Yang:2022qoy,Yang:2024idy,Niu:2024cfn,Ortega:2024zjx,Jia:2024ybo,Yang:2024iuc}. 
Within these approaches, the $\NNbar$ interaction has been described using phenomenological potentials~\cite{Cote:1982gr,Hippchen:1991rr,Mull:1991rs,Mull:1994gz,El-Bennich:2008ytt} and, more recently, chiral effective field theory (EFT)~\cite{Kang:2013uia,Dai:2017ont}.
Such a framework naturally connects the strong threshold enhancement in $e^+e^- \to N\bar N$ with possible two-step mechanisms of the type $e^+e^- \to N\bar N \to$ hadrons.
It also allows one to include effects that are essential for resolving the fine structure within a few MeV of threshold, such as the Coulomb interaction, the proton-neutron mass difference, and the coupled-channel $\ppbar$-$\nnbar$ dynamics~\cite{Milstein:2018orb,Jia:2024ybo}.
Alternative descriptions have also been put forward.
For example, vector-meson interference, in particular involving $\rho(1900)$, has been proposed to describe several hadronic channels in $e^+e^-$ annihilation~\cite{Lichard:2018enc}, while the near-threshold structures in $J/\psi$ decays have been interpreted in terms of pseudoscalar glueballs or excited $\eta^\prime$ states~\cite{Li:2005vd,Kochelev:2005vd,Hao:2005hu,Kochelev:2005tu,Huang:2005bc,Yu:2011ta,Gui:2019dtm,Wang:2020due}.

In this work, we study the near-threshold structures associated with $\NNbar$ production in $e^+e^-$ annihilation in the $J^{PC}=1^{--}$ sector. We examine whether the threshold behavior observed in both baryonic and non-baryonic final states can be understood within a unified description based on a single dynamical input for the coupled ${}^3S_1$-${}^3D_1$ $\NNbar$ system. The baryonic channels considered are $\ppbar$ and $\nnbar$, while the non-baryonic channels are $3(\pi^+\pi^-)$, $2(\pi^+\pi^-\pi^0)$, $2(\pi^+\pi^-)\pi^0$, $\omega\pi^+\pi^-\pi^0$, and $K^+K^-\pi^+\pi^-$. Chiral EFT provides a suitable framework for such an analysis. It has been successfully applied to the low-energy $NN$ interaction~\cite{Weinberg:1990rz, Weinberg:1991um,Ordonez:1993tn,Epelbaum:2008ga,Machleidt:2011zz} and has subsequently been extended to the $\NNbar$ sector~\cite{Chen:2011yu,Kang:2013uia,Dai:2017ont,Xiao:2024jmu}. Here we use the chiral EFT $\NNbar$ interaction constructed up to next-to-next-to-next-to-leading order (N${}^3$LO) in Ref.~\cite{Dai:2017ont}.

The paper is organized as follows. Section~\ref{sec:framework} describes the chiral EFT input for the $\NNbar$ interaction and the construction of the production amplitudes. The fit results are presented and discussed in
Sec.~\ref{sec:results}, and our conclusions are summarized in Sec.~\ref{sec:conclusion}.

\section{Framework}\label{sec:framework}

To describe the near-threshold structures in a unified manner, we proceed in three steps. First, the strong $\NNbar$ interaction is fixed by the chiral EFT description of the coupled ${}^3S_1$-${}^3D_1$ system. Second, keeping this interaction unchanged, we determine the electromagnetic production amplitudes from the $\ppbar$ and $\nnbar$ cross sections. Third, the non-baryonic channels are described by channel-dependent short-range production terms, parametrized as low-order polynomials in energy to account for slow varying background contributions, together with the same $\NNbar$ rescattering response.

\subsection{Chiral EFT for the $\NNbar$ interaction}

The chiral EFT description of the $\NNbar$ interaction up to N${}^3$LO was constructed in Ref.~\cite{Dai:2017ont}, where the low-energy constants were determined from the low-energy $\NNbar$ scattering information of Ref.~\cite{Zhou:2012ui}. In $e^+e^-$ annihilation through the timelike electromagnetic current, the relevant $\NNbar$ partial waves are the coupled ${}^3S_1$-${}^3D_1$ waves. For each isospin channel $I=0,1$, the scattering amplitude is obtained from the coupled-channel Lippmann--Schwinger equation (LSE)
\begin{align}
T^{I}_{L''L'}&(E;p'',p')
=
V^{I}_{L''L'}(p'',p')\notag\\
&+
\sum_{L}\int_0^\infty \frac{q^2dq}{(2\pi)^3}
\frac{
V^{I}_{L''L}(p'',q)\,
T^{I}_{LL'}(E;q,p')
}{
E-2E_N(q)+i\epsilon
},
\label{eq:lse}
\end{align}
where $E$ is the total energy, $p'$ and $p''$ are the incoming and outgoing momenta, and $E_N(q)=\sqrt{q^2+m_N^2}$, with $m_N$ the nucleon mass. The indices $L$, $L'$, and $L''$ run over the $S$- and $D$-wave components. The potential $V^I$ contains the long-range one- and two-pion-exchange contributions as well as short-range contact terms. The resulting amplitudes provide the strong $\NNbar$ rescattering input used in the production analysis below.

In this work, we use the $\NNbar$ interaction of Ref.~\cite{Dai:2017ont} directly, without refitting the low-energy constants. Our central results are based on the N${}^3$LO interaction with the regulator cutoff $R=0.9~{\rm fm}$. To estimate the residual uncertainty associated with the chiral expansion, we repeat the analysis with the lower-order interactions of Ref.~\cite{Dai:2017ont} and construct the truncation-uncertainty bands following the prescription used there. At each chiral order, the short-distance production parameters are refitted. Unless stated otherwise, the quoted results refer to the central N${}^3$LO result with $R=0.9~{\rm fm}$.

For the strong $\NNbar$ interaction of Ref.~\cite{Dai:2017ont}, one near-threshold pole is found in each isospin channel. The pole positions are
\begin{equation}
\begin{aligned}
E_{I=1}
&=
\left(2122\pm65\right)+i\left(30\pm57\right)\ {\rm MeV}
\quad \text{on RS$_{-}$},\\
E_{I=0}
&=
\left(1840\pm21\right)-i\left(80\pm6\right)\ {\rm MeV}
\quad~~\text{on RS$_{+}$}.
\end{aligned}
\label{eq:poles}
\end{equation}
Here, RS$_{+}$ and RS$_{-}$ denote the Riemann sheets on which the imaginary part of the on-shell $\NNbar$ momentum is positive and negative, respectively. Because the effective $\NNbar$ interaction contains an absorptive annihilation component, the poles acquire finite imaginary parts. Their proximity to the $\NNbar$ thresholds makes them relevant for the near-threshold line shapes in both $\NNbar$ and multi-hadron channels~\cite{Guo:2014iya,Dong:2020hxe}.

% The central pole positions in Eq.~\eqref{eq:poles} are obtained with the N${}^3$LO interaction at $R=0.9~{\rm fm}$. The quoted uncertainties estimate the chiral EFT truncation uncertainty following the prescription of Ref.~\cite{Dai:2017ont}, applied to the complex binding energy $E_{\rm pole}-2m_N$ obtained from the LO, NLO, N${}^2$LO, and N${}^3$LO interactions. This estimate reflects the convergence pattern of the chiral expansion and should not be interpreted as a complete uncertainty estimate of the pole positions.

The central pole positions in Eq.~\eqref{eq:poles} are obtained with the N${}^3$LO interaction at $R=0.9~{\rm fm}$. The quoted uncertainties estimate the chiral EFT truncation uncertainty following the prescription of Ref.~\cite{Dai:2017ont}, applied to the complex binding energy $E_{\rm pole}-2m_N$ obtained from the LO, NLO, N${}^2$LO, and N${}^3$LO interactions. For the $I=1$ pole, some lower-order results move to the adjacent RS$_{+}$ sheet, although the central N${}^3$LO pole lies on RS$_{-}$. Therefore this proximity to the sheet connection results in the comparatively large uncertainty in $\operatorname{Im} E_{I=1}$. The quoted uncertainty therefore characterizes the order-by-order stability of the pole trajectory, rather than a complete uncertainty estimate of the pole positions.

\subsection{Cross sections for $\ee\to N\bar N$}

The near-threshold enhancements observed in the $\ppbar$ and $\nnbar$ cross sections have attracted considerable attention. As discussed above, the strong $\NNbar$ final-state interaction generates near-threshold poles. Therefore, even energy-independent short-distance electromagnetic sources can lead to pronounced threshold structures once they are dressed by this final-state interaction. Since the chiral EFT description of the $\NNbar$ interaction is restricted to the low-energy region, we include experimental data only from the corresponding thresholds up to $E=2.0~{\rm GeV}$.

To account for isospin-breaking effects, including the $\ppbar$-$\nnbar$ threshold difference and the electromagnetic interaction, we work in the particle basis and label the $\ppbar$ and $\nnbar$ channels as channel 1 and channel 2, respectively. The strong potential in this basis is
\begin{align}
V
=
\frac{1}{2}
\begin{pmatrix}
V^{0}+V^{1} & V^{0}-V^{1} \\
V^{0}-V^{1} & V^{0}+V^{1}
\end{pmatrix},
\label{eq:stageb-particle-potential}
\end{align}
where the momentum and partial-wave labels have been suppressed for brevity. The physical proton and neutron masses are used in the corresponding particle-basis LSE.

We combine the partial-wave and particle-channel labels into a composite index,
\begin{equation}
    \alpha=(L,a),
\end{equation}
where $L=S,D$ and $a=p,n$ denotes the $\ppbar$ and $\nnbar$ channels, respectively. When needed, $L_\alpha$ and $a_\alpha$ denote the partial-wave and particle-channel components of $\alpha$. For a given total energy $E$ and
center-of-mass momentum $p'$, the full production amplitude $\mathcal A_\alpha$ is obtained by dressing the short-range electromagnetic source $\mathcal P_\alpha$ with the $\NNbar$ final-state interaction. In the
spirit of the distorted-wave Born approximation~\cite{Dai:2017fwx,Dai:2018tlc},
we write
\begin{align}
\mathcal{A}_{\alpha}(E,p')
&=
\mathcal P_{\alpha}(p')
\notag\\
&\quad+
\sum_{\beta}
\int_0^\infty \frac{q^2dq}{(2\pi)^3}
\frac{
T_{\alpha\beta}(E;p',q)\,
\mathcal P_{\beta}(q)
}{
E-2E_{\beta}(q)+i\epsilon
},
\label{eq:em-dressed-source}
\end{align}
where
$
E_{\beta}(q)=\sqrt{q^2+m_{\beta}^2},
$
with $m_{\beta}$ the nucleon mass in the particle channel specified by $\beta$.

The short-range sources are assumed to contain no near-threshold singularities and to vary only slowly with the total energy. Over the narrow energy region considered here, their energy dependence is therefore neglected, and they are parametrized as
\begin{align}
{\cal P}_{\alpha}(p')
=
c_{\alpha}\,f_{L_\alpha}(p').
\label{eq:stageb-source-isospin}
\end{align}
The regulator functions $f_L(p')$ suppress the high-momentum components of the source. We use the Gaussian form
\begin{align}
f_S(p')
&=
\exp\left(-\frac{p'^2}{\Lambda^2}\right),
\notag\\
f_D(p')
&=
\frac{p'^2}{\Lambda^2}
\exp\left(-\frac{p'^2}{\Lambda^2}\right).
\label{eq:stageb-source-regulator}
\end{align}
The cutoff is chosen as $\Lambda=0.45~{\rm GeV}$, which matches the momentum scale associated with the $R=0.9~{\rm fm}$ regulator used for the strong-interaction input in Ref.~\cite{Dai:2017ont}. The four coefficients $c_{(S,p)}$, $c_{(S,n)}$, $c_{(D,p)}$, and $c_{(D,n)}$ encode the short-distance physics of the electromagnetic production mechanism. Since the $1^{--}$ $\NNbar$ system can couple to many inelastic channels, these coefficients are allowed to be complex. After fixing an overall phase, the four complex coefficients contain seven real fit parameters.

For the electromagnetic interaction, we follow the treatment of Ref.~\cite{Dai:2017ont}. Only the Coulomb interaction in the charged channel is included, and it is implemented by means of the Vincent--Phatak (VP) method~\cite{Vincent:1974zz}. Other electromagnetic effects, such as the magnetic-moment interaction, are not included in the present calculation. This approximation is expected to have only a minor impact, as suggested by Ref.~\cite{Stoks:1990us}.

In the VP method, coordinate space is separated into short- and long-range regions by a separation radius $\mathcal R$, chosen sufficiently large that the strong interaction, including the long-range one-pion-exchange potential, is negligible for $r>\mathcal R$. The finite-range inner Coulomb interaction,
\begin{align}
    V_C^\mathcal R(r)= -\frac{\alpha}{r}\theta(\mathcal R-r),
\end{align}
is then added to the $\NNbar$ potential in the charged channel, with $\alpha=e^2/4\pi\simeq 1/137$ the electromagnetic fine-structure constant. After solving the coupled-channel scattering problem with this finite-range Coulomb interaction, the resulting finite-range production amplitudes are matched to the exact Coulomb asymptotic normalization through the VP matrix $N_{\alpha\beta}(E)$, defined in Appendix~\ref{app:VP}. The on-shell amplitudes entering the electromagnetic form factors are then
\begin{equation}
\widetilde{\cal A}_{\alpha}(s)
=
\sum_{\beta}
N_{\alpha\beta}(\sqrt{s})\,
{\cal A}_{\beta}(\sqrt{s},q_{\beta}),
\label{eq:coulomb-dressed-amplitudes}
\end{equation}
where $s=E^2$, and $q_\beta$ denotes the on-shell momentum in channel $\beta$. Although ${\cal A}_{\alpha}$ and $N_{\alpha\beta}$ separately depend on the separation radius $\mathcal R$, this dependence cancels in the final Coulomb-dressed amplitudes $\widetilde{\cal A}_{\alpha}$ when the VP matching is implemented properly. Further details are given in Appendix~\ref{app:VP}.

The dressed $S$- and $D$-wave production amplitudes are converted to the Sachs form factors through the standard projection from the ${}^3S_1$-${}^3D_1$ partial-wave basis to the transverse and longitudinal electromagnetic amplitudes,
\begin{align}
G_M^a(s)
&=
\frac{1}{3}
\left[
2\,\widetilde{\cal A}_{S,a}(s)
+
\sqrt{2}\,\widetilde{\cal A}_{D,a}(s)
\right],
\notag\\
G_E^a(s)
&=
\frac{\sqrt{s}}{3m_a}
\left[
\widetilde{\cal A}_{S,a}(s)
-
\sqrt{2}\,\widetilde{\cal A}_{D,a}(s)
\right].
\label{eq:GEGM-ASAD}
\end{align}
The cross section is evaluated as
\begin{align}
\sigma_{a}(s)
&=
\frac{4\pi\alpha^2\beta_a}{3s}
\left[
\left|G_M^a(s)\right|^2
+
\frac{2m_a^2}{s}
\left|G_E^a(s)\right|^2
\right],
\label{eq:nnbar-cross-section}
\end{align}
with $\beta_a=\sqrt{1-{4m_a^2}/{s}}$. In the present prescription, the Coulomb enhancement associated with the charged wave function is already included through Eq.~\eqref{eq:coulomb-dressed-amplitudes}. Therefore, no additional Sommerfeld factor is multiplied into Eq.~\eqref{eq:nnbar-cross-section}.

\subsection{Cross sections for the inelastic channels}

We next examine whether the same $\NNbar$ dynamics leaves visible imprints in non-baryonic final states. In what follows, these final states are referred to as inelastic channels. For an inelastic channel $h$, the cross section for $\ee\to h$ is written as
\begin{align}
\sigma_h(E)
=
\frac{\rho_h(E)
\left|{\cal A}_h(E)\right|^2}{2E^2} .
\label{eq:xsection-C}
\end{align}
The phase-space factor is fixed by kinematics and normalized at the $\ppbar$ threshold,
\begin{equation}
\rho_h(E)
=
\frac{\Phi_h(E)}{\Phi_h(2m_p)}.
\label{eq:stagec-phase-space}
\end{equation}
For multi-particle final states, intermediate resonances may contribute and complicate the evaluation of the phase space. In the present analysis, such substructures are not resolved explicitly. We therefore approximate $\Phi_h(E)$ by the pure multi-body phase space of channel $h$, neglecting final-state interactions among the produced hadrons.

The full production amplitude ${\cal A}_h(E)$ is written as the sum of a short-range production amplitude and a contribution from intermediate $\NNbar$ states,
\begin{align}
\mathcal A_{h}(E)
=&
\mathcal P_{h}(E)
\notag\\
&+
\sum_{\alpha}
\int_0^\infty \frac{q^2dq}{(2\pi)^3}
\frac{{\cal V}_{h\alpha}(E,q)\,\mathcal{A}_{\alpha}(E,q)
}{
E-2E_{\alpha}(q)+i\epsilon
}.
\label{eq:hadronic-master}
\end{align}
Here, ${\cal V}_{h \alpha }(E,q)$ describes the transition from the $\NNbar$ channel $\alpha$ to the final state $h$. Notice that $\mathcal{A}_{\alpha}(E,q)$ is the electromagnetic production amplitude of the intermediate $\NNbar$ state defined in Eq.~\eqref{eq:em-dressed-source}, instead of the VP normalized one in Eq.~\eqref{eq:coulomb-dressed-amplitudes}, because only off-shell $\NNbar$ amplitudes are needed in the integrand, which are not associated with asymptotic states.

The slowly varying direct-production term for each inelastic channel is parameterized as a real polynomial,
\begin{align}
\mathcal P_h(E)
&=
\sum_{j=0}^{j_{\rm max}}
b_{hj}
\left(
\frac{E^2-4m_p^2}{4m_p^2}
\right)^j .
\label{eq:stagec-background}
\end{align}
We have tested different values of $j_{\rm max}$ and find that $j_{\rm max}=2$ is sufficient to describe the data in the energy region considered here. The coefficients $b_{hj}$ encode the short-distance production mechanism of the inelastic channel $h$.

The transition amplitudes ${\cal V}_{h \alpha }$ are assumed to be slowly varying in energy over the narrow energy range considered here and are parameterized by complex constants $v_{h\alpha}$ multiplied by the same momentum regulators as in Eq.~\eqref{eq:stageb-source-regulator},
\begin{align}
{\cal V}_{h \alpha }(E,p')
=
v_{h\alpha} f_{L_\alpha}(p'),
\label{eq:stagec-transition}
\end{align}
To reduce the number of parameters, we keep only the $S$-wave transition amplitude, which is expected to dominate near threshold. We have checked that including the $D$-wave transition does not improve the fit significantly.

The allowed isospin components are fixed according to the isospin and $G$-parity quantum numbers of the final states. Among the five channels considered here, $2(\pi^+\pi^-)\pi^0$ and $\omega\pi^+\pi^-\pi^0$ are assigned to isospin-0, while $2(\pi^+\pi^-\pi^0)$ and $3(\pi^+\pi^-)$ are assigned to isospin-1. The channel $K^+K^-\pi^+\pi^-$ has both components. Equivalently, the particle-basis transition parameters satisfy $v_{h(L,p)}=v_{h(L,n)}$ for the isospin-0 channels and $v_{h(L,p)}=-v_{h(L,n)}$ for the isospin-1 channels. Therefore, in addition to the short-range production parameters in Eq.~\eqref{eq:stagec-background}, this step contains six independent complex $\NNbar\to h$ transition parameters as listed below in Table~\ref{tab:fit-parameters-hesse}.

\section{Results and discussion}\label{sec:results}

\begin{figure*}[h]
\centering
\includegraphics[width=0.92\textwidth]{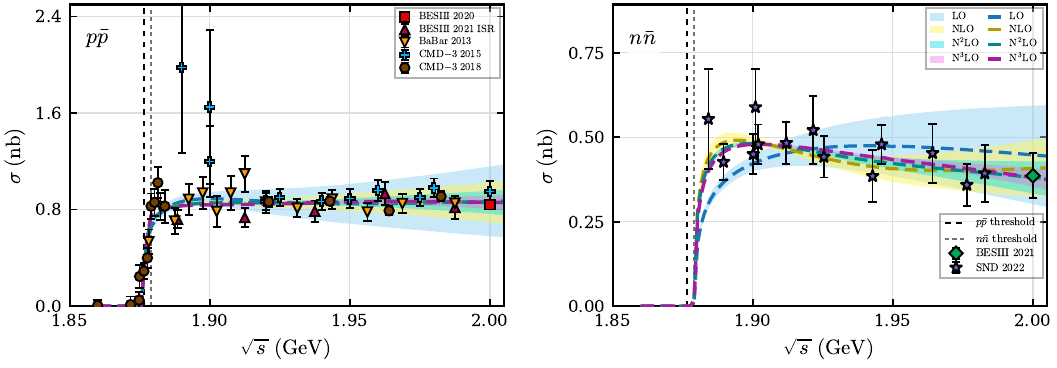}
\caption{Cross sections for $e^+e^- \to \ppbar$ and $e^+e^- \to \nnbar$ with the best-fit line shapes.
The curves show the results obtained with the $\NNbar$ interactions at different chiral orders, and the shaded bands denote the corresponding truncation uncertainties estimated following Ref.~\cite{Dai:2017ont}. The $\ppbar$ data are taken from BESIII 2020~\cite{BESIII:2019hdp}, BESIII 2021~\cite{BESIII:2021rqk}, BaBar 2013~\cite{BaBar:2013ves}, CMD-3 2015~\cite{CMD-3:2015fvi}, and CMD-3 2018~\cite{CMD-3:2018kql}. The $\nnbar$ data are taken from BESIII 2021~\cite{BESIII:2021tbq} and SND 2022~\cite{SND:2022wdb}. The fit is performed in the energy region from the respective threshold up to 2.0~GeV.}
\label{fig:stage2}
\end{figure*}
\begin{figure*}[h]
\centering
\includegraphics[width=0.92\textwidth]{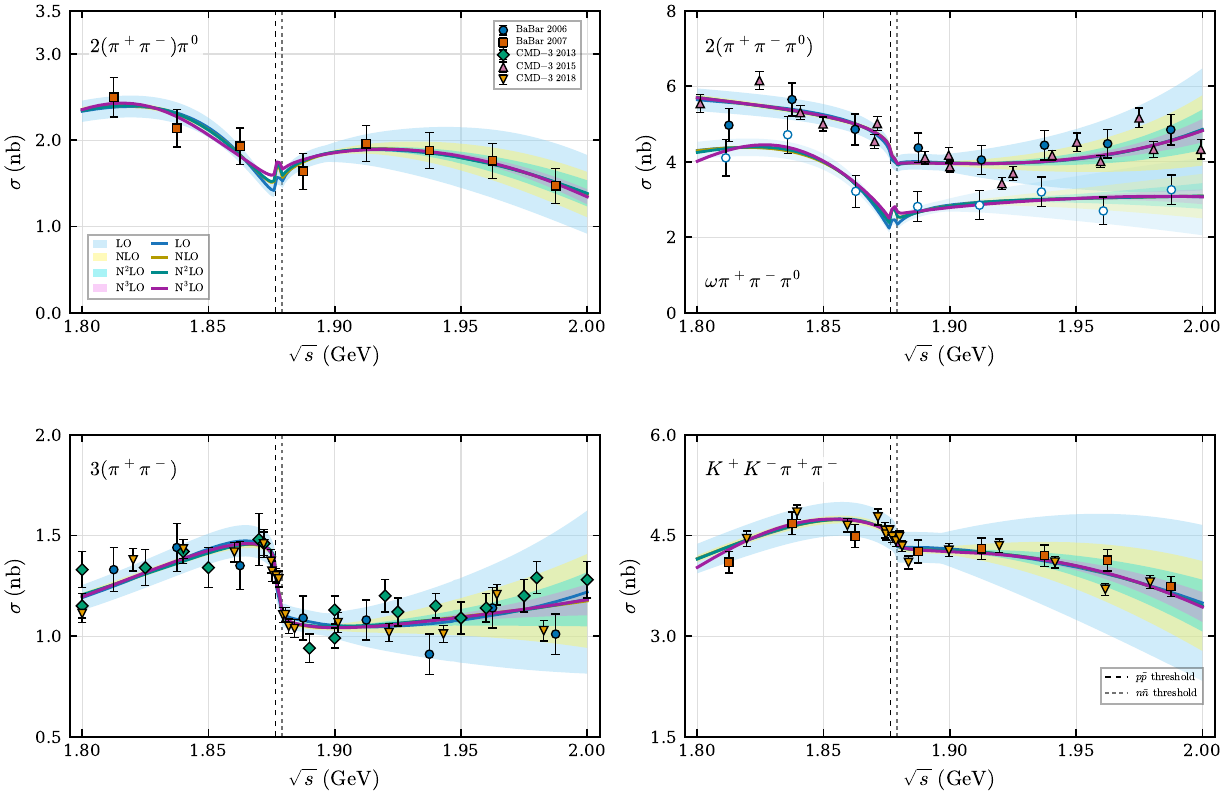}
\caption{
Cross sections for the inelastic channels with the best-fit line shapes. The curves show the full fits including the $\NNbar$ rescattering response, while the shaded bands indicate the truncation uncertainties estimated following Ref.~\cite{Dai:2017ont}. The data are taken from Refs.~\cite{BaBar:2006vzy,BaBar:2007ptr,CMD-3:2015fvi,CMD-3:2018kql,Lukin:2015vsa}.
}
\label{fig:stage3}
\end{figure*}

\subsection{Cross sections for $\ee\to\NNbar$}
\label{sec:results-A}
% Using the $\NNbar$ scattering amplitudes obtained from the chiral EFT potential, we fit the cross sections for $\ee\to N\bar N$, constructed according to Eq.~\eqref{eq:nnbar-cross-section}, to the available $\ppbar$ and $\nnbar$ data.
Using the $\NNbar$ scattering amplitudes obtained from the chiral EFT potential, we fit the $\ee\to\ppbar$ and $\ee\to\nnbar$ cross sections evaluated according to Eq.~\eqref{eq:nnbar-cross-section}. For the charged channel, we use the data from Refs.~\cite{BaBar:2013ves,CMD-3:2015fvi,CMD-3:2018kql,BESIII:2019hdp,BESIII:2021rqk}, while for the neutral channel we use those from Refs.~\cite{BESIII:2021tbq,SND:2022wdb}.
The data set contains 54 cross-section points in the energy range $2m_p\le \sqrt{s}\le 2.0~{\rm GeV}$, which covers the near-threshold region where the chiral EFT description of the $\NNbar$ interaction is expected to be applicable.

The fit contains seven real free parameters, corresponding to the four complex particle-basis source coefficients $c_{\alpha}$ defined in Eq.~\eqref{eq:stageb-source-isospin}, after fixing the overall phase by taking $c_{(S,p)}$ to be real and positive. The resulting best-fit values and uncertainties are listed in Table~\ref{tab:fit-parameters-B}. The uncertainties here quoted for these parameters are statistical fit errors only and should not be interpreted as EFT truncation uncertainties. In particular, the source coefficients absorb order-dependent changes in the $\NNbar$ final-state interaction. Consequently, the fitted values of the production parameters can vary substantially from one chiral order to another. The best-fit curves are shown in Fig.~\ref{fig:stage2}, where the shaded bands indicate the truncation uncertainties estimated from the order-by-order chiral EFT interactions following Ref.~\cite{Dai:2017ont}. Notice that the broadening of the bands at higher energies reflects the decreasing applicability of the chiral EFT description as the energy moves farther away from the $\NNbar$ thresholds. We obtain $\chi^2=45.9$ for 47 degrees of freedom, corresponding to $\chisqdof=1.0$, with individual contributions $\chi^2_{\ppbar}=41.2$ and $\chi^2_{\nnbar}=4.7$. Both the $\ppbar$ and $\nnbar$ cross sections are well reproduced in the fitted energy region.

These results demonstrate that the chiral EFT description of the $\NNbar$ interaction captures the main near-threshold enhancements in both the $\ppbar$ and $\nnbar$ channels. We note, however, that the CMD-3 points closest to the $\ppbar$ threshold require some care. As for all measured cross sections, the quoted Born cross sections are extracted from visible cross sections through experiment-specific corrections. This extraction becomes particularly sensitive in the immediate threshold region, where the cross section varies rapidly on the MeV scale and is affected by initial-state radiation, the beam-energy spread, and the reconstruction efficiency for slow antiprotons. Since our theory curves are evaluated at the Born level and do not implement the CMD-3 smearing and extraction procedure~\cite{CMD-3:2018kql}, we do not attempt a point-by-point description of these fine-scan points.

In particular, no additional narrow resonance is required to account for the rapid rise of the cross sections just above thresholds. This conclusion is supported by the fact that the strong $\NNbar$ input is fixed from low-energy $\NNbar$ scattering and is not adjusted to reproduce the threshold structures in the production data.

\begin{table}[t]
\caption{
Best-fit source parameters for the $\ee\to\NNbar$ cross sections.
}
\label{tab:fit-parameters-B}
\centering
\begin{ruledtabular}
\begin{tabular}{cc}
Parameter & {Value} \\
\hline
$c_{S,\ppbar}$ & $3.73(1.04)$\\
$c_{S,\nnbar}$ & $-2.08(1.17)+0.43(1.88)i$\\
$c_{D,\ppbar}$ & $0.43(0.18)-0.47(0.41)i$\\ 
$c_{D,\nnbar}$ & $0.72(0.46)-0.20(0.57)i$\\
\end{tabular}
\end{ruledtabular}
\end{table}

\subsection{Cross sections for the inelastic channels}

Having fixed the electromagnetic production amplitudes for $\NNbar$ in Sec.~\ref{sec:results-A}, we now turn to the inelastic channels, which are analyzed using Eq.~\eqref{eq:xsection-C}. The fit includes data for $e^+e^-$ annihilation into $2(\pi^+\pi^-)\pi^0$ from BaBar 2007~\cite{BaBar:2007ptr}, $2(\pi^+\pi^-\pi^0)$ from BaBar 2006~\cite{BaBar:2006vzy} and CMD-3 2015~\cite{Lukin:2015vsa}, $\omega\pi^+\pi^-\pi^0$ from BaBar 2006~\cite{BaBar:2006vzy}, $3(\pi^+\pi^-)$ from BaBar 2006~\cite{BaBar:2006vzy}, CMD-3 2015~\cite{CMD-3:2015fvi}, and CMD-3 2018~\cite{CMD-3:2018kql}, and $K^+K^-\pi^+\pi^-$ from BaBar 2007~\cite{BaBar:2007ptr} and CMD-3 2018~\cite{CMD-3:2018kql}.

The resulting line shapes are shown in Fig.~\ref{fig:stage3}, and the fitted parameters are listed in Table~\ref{tab:fit-parameters-hesse}. The best fit gives a total $\chi^2=140.8$ for 110 data points and 83 degrees of freedom, corresponding to $\chisqdof=1.7$.  The fit quality for each channel is given in the second column of Table~\ref{tab:fit-parameters-hesse}. Overall, the fitted line shapes reproduce the main near-threshold features of the data.

\begin{table*}[t]
\caption{
Fit parameters for the inelastic channels in the particle basis from the N3LO Stage-3 fit. The parameters $b_{hj}$ are the coefficients of the short-range production term defined in Eq.~\eqref{eq:stagec-background}. The parameters $v_{h}^{p}\equiv v_{h(S,p)}$ and $v_{h}^{n}\equiv v_{h(S,n)}$ are the $\ppbar$ and $\nnbar$ $S$-wave components of the $\NNbar\to h$ transitions defined in Eq.~\eqref{eq:stagec-transition}. In the Value column, the entries for $b_{hj}$ are quoted in units of $10^{-3}$, while the entries for $v_h^{p,n}$ are quoted in units of $\mathrm{GeV}^{-2}$.
}
\label{tab:fit-parameters-hesse}
\centering
\setlength{\tabcolsep}{3pt}
\begin{ruledtabular}
\begin{tabular}{lccc}
 Channel ($h$) &$\chi_h^2/N_h$ & Parameter &  Value \\
\hline
\multirow{4}{*}{$2(\pi^+\pi^-)\pi^0$} &\multirow{4}{*}{0.17}& $b_{h0}$ & $6.05(0.37)$\\
 && $b_{h1}$ & $-4.22(11.17)$\\
 && $b_{h2}$ & $-74.45(85.55)$\\
 && $v_{h}^{p}=v_{h}^{n}$ & $0.60(1.70)+3.25(2.04)i$\\
\hline
\multirow{4}{*}{$2(\pi^+\pi^-\pi^0)$} &\multirow{4}{*}{2.15}& $b_{h0}$ & $9.00(0.71)$\\
 && $b_{h1}$ & $-26.21(9.22)$\\
 && $b_{h2}$ & $92.83(75.75)$\\
 && $v_{h}^{p}=-v_{h}^{n}$ & $-4.11(3.41)+0.42(1.49)i$\\
\hline
\multirow{4}{*}{$3(\pi^+\pi^-)$} &\multirow{4}{*}{1.25}& $b_{h0}$ & $3.99(0.15)$\\
 && $b_{h1}$ & $-4.65(0.98)$\\
 && $b_{h2}$ & $-0.08(10.62)$\\
 && $v_{h}^{p}=-v_{h}^{n}$ & $-3.81(0.38)+1.49(0.29)i$\\
\hline
\multirow{5}{*}{$K^+K^-\pi^+\pi^-$} &\multirow{5}{*}{1.13}& $b_{h0}$ & $8.19(3.93)$\\
 && $b_{h1}$ & $-11.90(36.98)$\\
 && $b_{h2}$ & $-38.56(157.36)$\\
 && $v_{h}^{p}$ & $0.43(7.91)+2.28(13.08)i$\\
 && $v_{h}^{n}$ & $1.93(1.12)-0.77(0.42)i$\\
\hline
\multirow{4}{*}{$\omega\pi^+\pi^-\pi^0$} &\multirow{4}{*}{0.30}& $b_{h0}$ & $7.03(0.59)$\\
 && $b_{h1}$ & $-3.65(17.08)$\\
 && $b_{h2}$ & $-28.59(127.33)$\\
 && $v_{h}^{p}=v_{h}^{n}$ & $4.13(2.65)+4.17(3.51)i$\\
\end{tabular}
\end{ruledtabular}
\end{table*}

\section{Conclusion}\label{sec:conclusion}

We have studied the near-threshold structures observed in $\ee$ annihilation into $\NNbar$ and multi-hadron final states within a common $\NNbar$ final-state-interaction framework. The analysis proceeds in three steps. First, the strong $\NNbar$ interaction is taken from the chiral EFT description of the coupled ${}^3S_1$-${}^3D_1$ system, constrained by low-energy $\NNbar$ scattering data. Second, the $\ppbar$ and $\nnbar$ cross sections are used to determine the short-distance electromagnetic production of $\NNbar$, dressed by this fixed final-state interaction. Third, several inelastic channels are described by channel-dependent production terms together with the same universal $\NNbar$ rescattering response.

With this setup, the near-threshold behaviors of the $\ppbar$ and $\nnbar$ cross sections, as well as the structures observed in several inelastic channels, can be described consistently by the same coupled-channel $\NNbar$ dynamics. The nearby poles generated by the strong $\NNbar$ interaction play an essential role in shaping the line shapes around the $\ppbar$ and $\nnbar$ thresholds. 

These results indicate that the prominent near-threshold structures need not be attributed to separate narrow resonances introduced independently in different channels. Rather, they arise naturally from a common $\NNbar$ final-state interaction, showing that a naive Breit--Wigner interpretation may be misleading in the near-threshold region if the underlying rescattering dynamics is not taken into account. This interpretation is further supported by the truncation-uncertainty analysis based on the order-by-order chiral EFT interactions of Ref.~\cite{Dai:2017ont}. After refitting the short-distance production parameters at each chiral order, the resulting line shapes remain close to the central N${}^3$LO results, indicating that the qualitative conclusion is stable against the estimated truncation uncertainty.

\subsection*{Acknowledgments}
This work is supported in part by
Deutsche Forschungsgemeinschaft (DFG) under Grant No. 525056915 and under Germany's Excellence Strategy -- EXC 3107 -- Project-ID~533766364,  by the CAS President's International Fellowship Initiative (PIFI) under Grant Nos. 2025PD0022 and by  the European
Research Council (ERC) under the European Union's Horizon 2020 research
and innovation programme (EXOTIC, grant agreement No. 101018170).

\appendix
\section{Vincent--Phatak matching for the Coulomb interaction}
\label{app:VP}

Here, we describe the Coulomb treatment used in Eq.~\eqref{eq:coulomb-dressed-amplitudes}. We use the same composite particle-basis index as in the main text, $\alpha=(L,a)$,
so that the coupled-channel space is spanned by
\begin{equation}
   \left\{ p\bar p\,{}^3S_1,\,
     p\bar p\,{}^3D_1,\,
     n\bar n\,{}^3S_1,\,
     n\bar n\,{}^3D_1\right\}.
\end{equation}
The Coulomb interaction acts only in the charged $p\bar p$ channel and is diagonal in $L$.

The VP method~\cite{Vincent:1974zz} separates the interaction region into an inner region, $r<\mathcal R$, and an outer region, $r>\mathcal R$. The separation radius $\mathcal R$ is chosen sufficiently large such that the strong interaction is negligible outside this radius. In the inner region one solves the coupled-channel problem with the finite-range potential
\begin{align}
    V_\mathcal R(r)&
    =
    V_N(r)+
    {\rm diag}\left(V_C(r)\theta(\mathcal R-r),0\right),\label{eq:VR}
\end{align}
where $V_N$ denotes the strong $\NNbar$ potential in coordinate space and
\begin{equation}
    V_C(r)=-\frac{\alpha}{r}
\end{equation}
is the Coulomb potential with $\alpha = e^2/4\pi\simeq 1/137$ the fine-structure constant.
The corresponding finite-range scattering amplitude is denoted by $T^R_{\alpha\beta}$.

The inner wave-function $\mathbf U(r)$ used in the matching is a matrix with the first index denoting the channel components of a given state and the second index denoting the unit incoming basis channel. 
 In component form,
\begin{align}
    U_{\alpha\beta}(E&,r)
    =
    \delta_{\alpha\beta}\hat j_{L_\alpha}(q_{\alpha} r)
    \notag\\
    &\quad+
    \int_0^\infty \frac{q^2dq}{(2\pi)^3}\,
    \frac{
    \hat j_{L_\alpha}(qr)\,
    T^R_{\alpha\beta}(q,q_{\beta};E)}
    {E-2E_{\alpha}(q)+i\epsilon}
    \,\frac{q_{\beta}}{q}.
    \label{eq:wave-from-half-onshell-coupled}
\end{align}
Here, $\hat j_L(x)=xj_L(x)$ is the Riccati--Bessel function, and $q_{\alpha}$ is the on-shell momentum of channel $\alpha$ for the given $E$. We denote the quantity $R_{\alpha\beta}(E,r)$ as the ordinary radial wave function, and our reduced wave matrix is defined as $U_{\alpha\beta}(E,r)=q_\beta r\,R_{\alpha\beta}(E,r)$. Therefore the factor $q_\beta/q$ in Eq.~\eqref{eq:wave-from-half-onshell-coupled} follows solely from this reduced-wave convention and is not an additional dynamical factor.

The inner solution obtained in this way is not yet the physical Coulomb wave function, because the long-range Coulomb tail for $r>\mathcal R$ has been removed. To restore it, one matches the logarithmic derivative of the inner wave function to the exact outer Coulomb solution at $r=\mathcal R$. The logarithmic derivative of the inner solution is
\begin{equation}
    \mathbf L(\mathcal R)
    =
    \mathbf U'(\mathcal R)\mathbf U^{-1}(\mathcal R).
    \label{eq:VP-log-derivative}
\end{equation}

In the outer region the long-range interaction is diagonal in the particle-basis channel space. More precisely, the charged $p\bar p$ components feel the Coulomb tail, while the neutral $n\bar n$ components are free. We therefore introduce diagonal matrices $\mathbf F_C$ and $\mathbf G_C$ by
\begin{align}
    [\mathbf F_C(E,r)]_{\alpha\beta}
    &=
    \delta_{\alpha\beta} f^C_\alpha(E,r),
    \qquad \notag\\
    [\mathbf G_C(E,r)]_{\alpha\beta}
    &=
    \delta_{\alpha\beta} g^C_\alpha(E,r),
\end{align}
where, for $\alpha=(L,a)$, we have
\begin{align}
     f^C_{(L,p)}(E,r)&=F_L(\eta_p,q_p r),\notag\\
    g^C_{(L,p)}(E,r)&=G_L(\eta_p,q_p r),\notag\\
    f^C_{(L,n)}(E,r)&=\hat j_L(q_n r),\notag\\
    g^C_{(L,n)}(E,r)&=\hat g_L(q_n r).
\end{align}
where $\eta_p=-(\alpha m_p)/(2q_p)$. Notice that $F_L$ and $G_L$ are the regular and irregular Coulomb Riccati functions, respectively, while $\hat j_L(x)=xj_L(x)$ and $\hat g_L(x)$ are the corresponding free regular and irregular Riccati--Bessel functions. More precisely, we have $F_L(0,x)=\hat j_L(x)$ and $G_L(0,x)=\hat g_L(x)$.

With these definitions, the matrix of outer solutions satisfying unit
incoming-wave boundary conditions can be written as
\begin{align}
    \bm\Omega_C(E,r)&= [\mathbf F_C(E,r)+\mathbf G_C(E,r)\mathbf K_C(E)] \notag\\
    &\times
    [\mathbf 1-i\mathbf K_C(E)]^{-1}
    \bm\Phi_C(E).
\end{align}
Each column of $\bm\Omega_C$ corresponds to one incoming basis channel. The matrix $\mathbf K_C$ is a full coupled-channel matrix, while $\mathbf F_C$, $\mathbf G_C$, and $\bm\Phi_C$ are diagonal because the outer Coulomb/free Hamiltonian does not mix channels. The Coulomb phase matrix is
\begin{align} \bm\Phi_C(E)=\operatorname{diag}\left(e^{i\sigma_0},e^{i\sigma_2},1,1\right).
\end{align}

Imposing equality of the logarithmic derivatives at separation $\mathcal R$:
\begin{equation}
    \mathbf L_C(\mathcal R)
    =
    \bm\Omega_C^{\prime}(\mathcal R)
    \bm\Omega_C^{-1}(\mathcal R),
\end{equation}
gives the coupled-channel $K$-matrix
\begin{align}
K& =
\left[ \mathbf L_C(\mathcal R) \mathbf G_C(\mathcal R)-\mathbf G_C'(\mathcal R) \right]^{-1}\notag\\
&\times\left[ \mathbf F_C'(\mathcal R)-L_C(\mathcal R)\mathbf F_C(\mathcal R) \right].
\end{align}
This expression is the coupled-channel generalization of the usual single-channel VP matching condition.

Matching the inner wave function to the normalized unit-incoming wave function in the outer region, we define the VP normalized matrix $\mathbf N$ as
\begin{equation}
    \mathbf U(\mathcal R)\mathbf N^T(E)=\bm\Omega_C(\mathcal R),
    \label{eq:VP-N-definition}
\end{equation}
so that
\begin{equation}
  \mathbf N(E)=\left[\mathbf U^{-1}(\mathcal R)\bm\Omega_C(\mathcal R)\right]^T.
\end{equation}
The production amplitude then reads
\begin{equation}
    \widetilde{\bm{\mathcal{A}}}(E)
    =
    \mathbf N(E)\bm{\mathcal{A}}(E),
    \label{eq:VP-relative-normalization}
\end{equation}
which is equivalent with Eq.~\eqref{eq:coulomb-dressed-amplitudes}.
\bibliography{ref}
\end{document}